\documentclass[11pt,letterpaper]{JHEP3}
\usepackage{cite} 
\input epsf.tex

\newcommand{\beq}{\begin{eqnarray}}
\newcommand{\eeq}{\end{eqnarray}}

\def\ltap{\ \raise.3ex\hbox{$<$\kern-.75em\lower1ex\hbox{$\sim$}}\ }
\def\gtap{\ \raise.3ex\hbox{$>$\kern-.75em\lower1ex\hbox{$\sim$}}\ }
\def\CO{{\cal O}}

\def\CO{{\cal O}}

\def\mpl{M_{\rm Pl}}

\def\eg{{\it e.g.}}
\def\ie{{\it i.e.}}

\def\be{\begin{equation}}
\def\ee{\end{equation}}
\def\bea{\begin{eqnarray}}
\def\eea{\end{eqnarray}}

\newcommand{\gsim}{ \mathop{}_{\textstyle \sim}^{\textstyle >} }
\newcommand{\lsim}{ \mathop{}_{\textstyle \sim}^{\textstyle <} }
\newcommand{\vev}[1]{ \left\langle {#1} \right\rangle }

\newcommand{\ev}{{\rm eV}}
\newcommand{\kev}{{\rm keV}}

\newcommand{\gev}{{\rm GeV}}

\newcommand{\mev}{{\rm MeV}}

\newcommand{\mnu}{\ensuremath{m_\nu}}
\newcommand{\mlr}{\ensuremath{m_{lr}}}
\newcommand{\acc}{\ensuremath{{\cal A}}}

\title
{Dark Energy from Mass Varying Neutrinos}
\author{Rob Fardon, Ann E. Nelson and Neal Weiner \\ 
Department of Physics, Box 1560, University of Washington,\\
           Seattle, WA 98195-1560, USA
}
\preprint{\today \\ }
\abstract{We show that mass varying neutrinos (MaVaNs) can behave as a negative pressure fluid which could be the origin of the cosmic acceleration. We derive a model independent relation between the neutrino mass  and the equation of state parameter of the neutrino dark energy, which is applicable for general theories of mass varying particles. The neutrino mass depends on the local neutrino density and  the observed neutrino mass can exceed the cosmological bound on a constant neutrino mass. We discuss microscopic realizations of the MaVaN acceleration scenario, which involve a sterile neutrino. We consider naturalness constraints for mass varying particles, and find that both \ev\  cutoffs and \ev\ mass particles are needed to avoid fine-tuning. These considerations give a  (current) mass of order an eV for the sterile neutrino in microscopic realizations, which could be  detectable at MiniBooNE. Because the sterile neutrino was much heavier at earlier times, constraints from big bang nucleosynthesis on additional states are not problematic. We consider regions of high neutrino density and find that the  most likely place today to find neutrino masses which are significantly different from the neutrino masses in our solar system   is in  a supernova. The possibility of different neutrino mass in different regions of the galaxy and the local group could be significant for Z-burst models of ultra-high energy cosmic rays. We also consider the cosmology  of and the constraints on the ``acceleron'', the  scalar field  which is responsible for the varying neutrino mass, and  briefly discuss neutrino density dependent variations in other constants, such as the fine structure constant.
}

\begin{document}

\section{Introduction}
\label{intro}

In recent years, we have learned much about the composition of the universe. Motivated initially by the evidence from supernovae that the universe is accelerating \cite{Riess:1998cb,Perlmutter:1998np}, and with recent data on the cosmic microwave background radiation (CMBR) \cite{deBernardis:2000gy,Stompor:2001xf,Bennett:2003bz}, the $\Lambda$CDM cosmology has become firmly established  as the benchmark against which all other cosmologies must be compared. The need for a mysterious new ``dark energy'' component to the universe has pointed to the existence of new physics beyond the standard model (BSM).

At almost the same time, the field of neutrino masses has undergone a revolution. There is the compelling evidence for atmospheric neutrino oscillations from the SuperKamiokande experiment \cite{Fukuda:1998mi}, consistent with earlier evidence from the IMB experiment \cite{Casper:1991ac,Becker-Szendy:1992hq} and Kamiokande \cite{Hirata:1992ku,Fukuda:1994mc}, and recently confirmed by K2K \cite{Walter:2002wh,Wilkes:2002mb,Miura:2002sb}. The SNO \cite{Ahmed:2003kj} and KamLAND \cite{Eguchi:2002dm} experiments have confirmed the neutrino oscillation interpretation of the solar neutrino deficit observed at other experiments \cite{Fukuda:1996sz,Hampel:1998xg,Altmann:2000ft,Abdurashitov:1994bc,Cleveland:1998nv}. The LSND result \cite{Athanassopoulos:1998pv} remains a puzzling anomaly, whose interpretation via neutrino oscillations will be tested at the upcoming MiniBooNE experiment \cite{McGregor:2003ds}. To add neutrino masses into the standard model, one must invoke either the existence of non-renormalizable operators, or light, sterile (singlets under the standard model gauge group) fermions, either of which point to the existence of BSM physics.

While both of these developments are tremendously exciting, giving the second and third direct pieces of evidence for new physics (the first being the large body of evidence for non-baryonic dark matter), they are both unsatisfying. Models of dark energy are  difficult to impossible to  test. Some give precisely a small cosmological constant as a result, others predict a non-standard equation of state, but little else to distinguish them. A notable exception has measurable long-distance modifications of gravity as a consequence \cite{Dvali:2000hr}, but a theoretical debate still rages around this model \cite{Luty:2003vm,Rubakov:2003zb}. Effective field theory considerations motivate modifications of gravity at the sub-millimeter scale \cite{Sundrum:1997js,Sundrum:2003jq}, but it is not obvious that we will be able to discern the nature of the dark energy, even in the presence of such measurements.

The situation for neutrino masses is not much better. Although the existence of neutrino mass is very exciting, it is not {\it a priori} clear how much we can learn  about fundamental physics from the neutrino masses. Majorana neutrino masses  can arise from a dimension five operator, whose origin can be from physics at an inaccessibly  high scale,  as high as $10^{15}\gev$. Although there are exciting ideas that relate neutrino masses to supersymmetry breaking \cite{Benakli:1997iu,Dvali:1998qy,Langacker:1998ut,Arkani-Hamed:2000bq,Arkani-Hamed:2000kj}, or the size of large extra dimensions \cite{Dienes:1998sb,Arkani-Hamed:1998vp}, the most popular seesaw models \cite{seesaw1,seesaw2} seem impossible to test directly.

\subsection{Coincidences and Damn Coincidences}
\label{damn}
The problem of the dark energy is especially puzzling because of what has become known as the ``cosmic coincidence problem.'' Namely, at the present time, $\rho_{CDM}/\rho_\Lambda \sim 1/3$, although this ratio changes as a function of cosmic scale factor as $1/a^3$. Hence, although these energy densities vary in dramatically different fashion over the history of the universe, we find that they are the same order at the present day.

This coincidence has been the focus of a great deal of work: namely how to
arrange for the dark matter and dark energy densities to be
comparable. Solutions to this are difficult in that we actually know a great
deal about the cosmological behavior of dark matter. Results from CMB let us
know that at $z\simeq 1100$ the universe was dominated by dark matter, and given
the observed density of dark matter, this is consistent with a fluid redshifting
with equation of state $w=0$. While we know little about the dark energy at
early times, we know that it is presently redshifting with equation of state
$w\approx -1$ \cite{Tonry:2003zg}. Thus it either has had $w\approx -1$ for a long
(cosmological) time, in which case there is a coincidence problem, or it tracked
dark matter and only recently switched to redshift slowly, in which case one has
a ``why-now?'' problem. In a sense, the
cosmic coincidence between dark matter and dark energy might be more properly
phrased as to why the dark energy {\em now} should be precisely three hundred
million times smaller than the dark matter energy density at recombination.

On top of this, we are in fact confronted with a {\em number} of coincidences.
There are many different components of the
energy density of the universe: dark matter, baryons, radiation, neutrinos as
well as dark energy. The first four all redshift very differently from with the
dark energy, and yet we have the remarkable coincidence that
they all have been equal to the dark energy within a
redshift of a few, shown in figure \ref{fig:cosmdensities}.  
\FIGURE[t]{
 \centerline{\epsfxsize=4.5 in \epsfbox{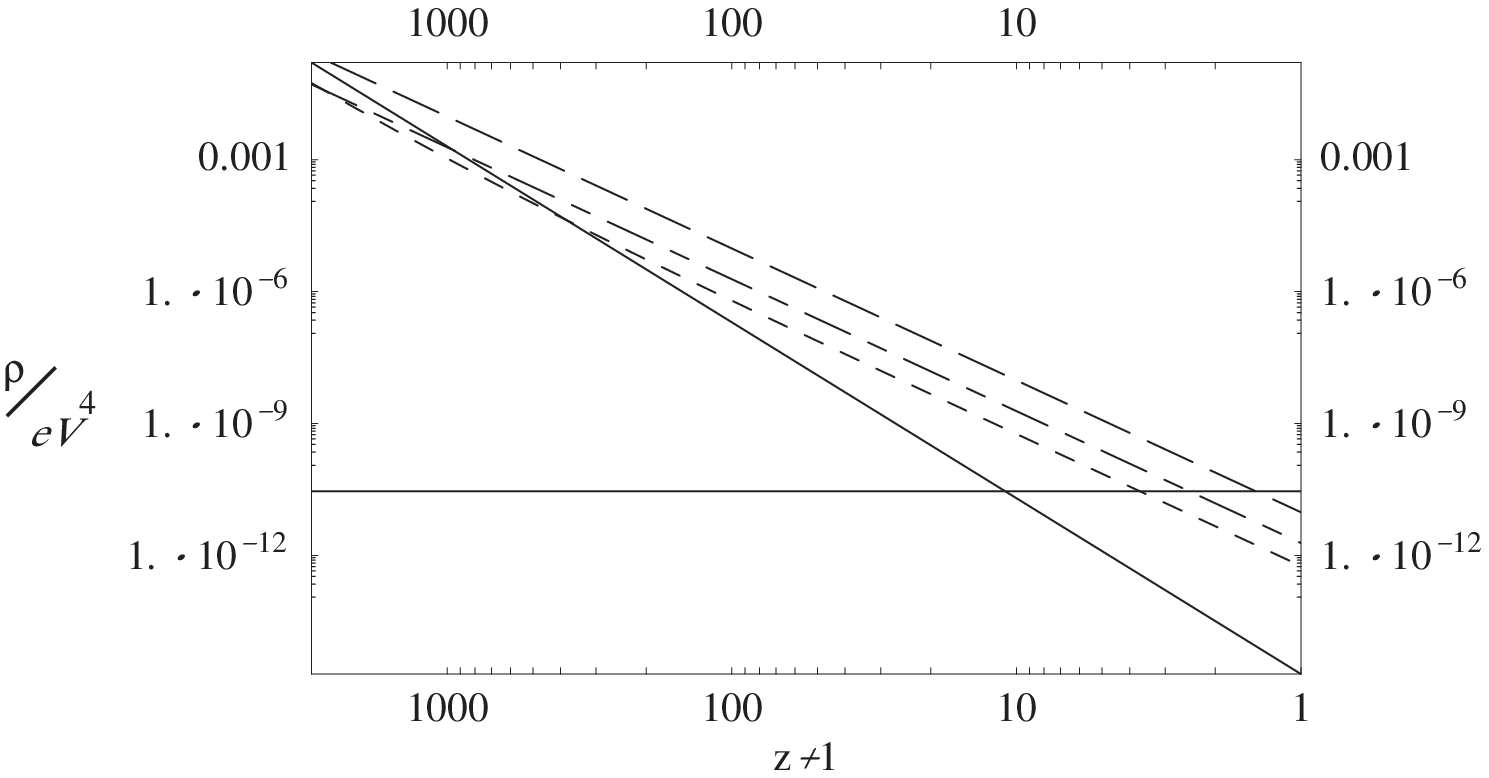}  }
\noindent
\caption{Cosmological energy densities as a function of redshift in the $\Lambda$CDM model. The different lines are: long dashed - CDM; long-short dashed - baryons; short-dashed - neutrinos; solid sloped - radiation; solid horizontal - cosmological constant.}
\label{fig:cosmdensities}
}
Is it possible that another of these other coincidences is not a coincidence?
While dark energy cannot reasonably track dark matter since recombination, is it
possible that dark energy has tracked one of these other components?
We have already discussed the difficulties in
explaining the dark matter coincidence. The baryon number of the universe,
constrained considerably by WMAP, is already probed at an even earlier era via
the success of big bang nucleosynthesis (BBN), meaning that dark energy should
not have been tracking baryons, unless it has the ``why-now?'' late time
transition of its equation of state.
Photon radiation is just as problematic.

Neutrinos, however, are an entirely different story. Currently,  the neutrino
energy density is quite uncertain. Large scale surveys and WMAP together
constrain $\Omega_\nu \lsim .02$. Terrestrial measurements of neutrino mass
indicate $\Omega_\nu > 7 \times 10^{-4}$. BBN implies that roughly three species
of neutrino were relativistic at BBN, which leads to the mild conclusion that
neutrino masses  are below $O(\mev)$. In reality, very little is known about the
cosmological behavior of neutrinos and the neutrino energy density. 

In this paper, we will propose that the coincidence $\rho_\nu \sim \rho_\Lambda$
(within a factor of $10^3$) is not a coincidence at all, but in fact a
relationship which holds over a large portion of the history of the
universe. This is possible if the mass of the neutrino is not a fixed parameter,
but a dynamical quantity, with an associated  potential, in analog with variable
mass dark matter\cite{Anderson:1997un}. The potential will have a zero energy
minimum at a value of $m_\nu$ which is larger than its present value. However,
the presence of a (cosmologically) uniform neutrino background will lead to an
effective potential which prevents $m_\nu$ from becoming too large, leaving a
homogeneous negative pressure fluid in the universe (\ie, dark energy). 

Variable mass particles have been previously considered in many contexts, with
attention recently revived in light of the current cosmological data. Early
consideration of such ideas for dark matter with varying mass appeared in
refs.~\cite{Casas:1992ky,Garcia-Bellido:1993de} in the context of scalar-tensor
gravity. Variable mass particles were proposed to solve the age of the universe
problem in ref.~\cite{Anderson:1997un}, and more recently as a solution to the
problem of the coincidence of dark matter and dark energy densities
\cite{Comelli:2003cv}. Phenomenological problems with this have been studied in
\cite{Farrar:2003uw,Hoffman:2003ru,Franca:2003zg}. In the context of neutrinos,
ref.~\cite{Kawasaki:1992gn} considered a Yukawa-type coupling to an extremely
light scalar field, leading to a different evolution of structure, and noted
that the neutrino mass would be density dependent, although  heavier rather than
lighter at earlier times. Ref.~\cite{Singh:1994nt}  argued that  dark energy in a desirable range  could be linked to    a scalar field coupled to  neutrinos with a desirable mass. Refs.~\cite{Stephenson:1997cz,Stephenson:1998qj}
considered a neutrino Yukawa interaction as a source for ``neutrino clouds'' which might
seed star formation, and change the mass of neutrinos in their
vicinity.  Ref.~\cite{Hung:2000yg} considered coupling a sterile neutrino to a slowly rolling, Hubble mass scalar field which was responsible for dark energy, but did not consider the impact of the cosmic neutrino background on the potential. Ref.~\cite{Gu:2003er} considered that varying neutrino mass  might have an impact on electroweak baryogenesis.  

In this  paper we consider the possibility that the  neutrino mass arises from
an interaction with a scalar field, the ``acceleron'', whose effective potential
changes as a function of the neutrino  density, and  consider the conditions
under which such a new ``dark force'', felt by neutrinos,  can explain the
observed acceleration of the universe. 
In section \ref{sec:MaVaNs} we will introduce the formalism and general features
for mass varying neutrinos (MaVaNs), focusing on model independent
relationships. We find a relationship between the neutrino mass and density, the
neutrino dark energy density, and the equation of state of  neutrino dark energy
which will hold for any theory of neutrinos of varying mass. We  discuss the
effects of varying  neutrino mass, specifying where we expect significant
variations both in time and space, and reconsidering the  traditional neutrino
mass constraints as they apply to variable masses. We find that some neutrino
mass bounds, such as those from large scale structure formation, do not
apply. In section \ref{sec:models} we  consider the details of particular
microscopic realizations. Radiative stability of the scalar potential uniquely
chooses theories with light particles and  a cutoff for radiative effects at low
energies (\ie, sub-eV). We argue that  variable mass WIMPs are  too heavy to be
significant contributors to the dark energy, phenomenological considerations
aside. A simple realization of MaVaNs arises from integrating out a standard
model singlet fermion whose mass depends on the expectation value of the
acceleron field. We note that traditional cosmological constraints from BBN on
sterile neutrinos do not apply here, and that naturalness suggests that
MiniBooNE will be able to detect such a sterile neutrino.  In section
\ref{sec:pheno} we consider phenomenological implications of MaVaNs, including
effects at terrestrial neutrino experiments, as well as in solar oscillations and
supernovae.

\section{MaVaNs}
\label{sec:MaVaNs}
MaVaNs have a number of significant phenomenological
consequences, and a great deal of this scenario can be understood in a
completely model independent way. To begin, we consider the neutrino mass
$m_\nu$ to be a dynamical field. Note that although for some purposes one should
study  \mnu\ as a  function of a canonically normalized scalar field $\acc$, as
long as $\partial \mnu /\partial \acc$ does not vanish, for the purpose of
analyzing the minimal energy density it suffices to write the scalar potential
as a function of \mnu. For simplicity, we only consider  a single nonvanishing
neutrino mass, although generalization to more than one  mass is
straightforward. 

In general, the contribution of a neutrino background to the energy density is given by
\beq
\delta V = \int \frac{d^3 k}{(2\pi)^3} {\sqrt{k^2 + \mnu^2}} f(k)\ ,
\label{eq:deltav}
\eeq
where $f(k)$ is the sum of the neutrino and anti neutrino occupation numbers for momentum $k$. 

The dependence of this  energy density on a change in the neutrino mass is given by 
\beq
\frac{\partial \delta V }{\partial m_\nu}\equiv s_\nu = \int\frac{d^3 k}{(2\pi)^3}  \frac{m_\nu}{\sqrt{k^2 + \mnu^2}} f(k)\ .
\eeq
The quantity $s_\nu$ , which we refer to as the ``scalar neutrino density'', turns out to be important for MaVaN cosmology and phenomenology as  the source term for a scalar field is proportional to $s_\nu$. In the nonrelativistic limit $s_\nu=n_\nu$, the total density of neutrinos and anti neutrinos.

For the cosmological Big Bang remnant neutrinos we have a contribution to the effective  potential 
\beq
\delta V = \int \frac{d^3 k}{(2\pi)^3} {2 \sqrt{k^2 + \mnu^2}\over{1+\exp(\beta_0\sqrt{m_0^2+k^2z^2})}},
\eeq
where $\beta_0 = 1/kT_0$ is the temperature at freeze out, $m_0$ is the neutrino mass at freeze out, and $z$ is the redshift at freeze out. 
In the nonrelativistic limit, the  leading \mnu\ dependent term in the energy density from the neutrino background is simply $n_\nu \mnu$. 
Therefore, in this limit the dependence of the energy density on \mnu\ is given by the effective potential
\beq
V(m_\nu) = m_\nu n_\nu + V_0(m_\nu)\ ,
\eeq
where $V_0(m_\nu)$ is the scalar potential. The presence of thermal background neutrinos acts as a source which will drive \mnu\ to small values. On the other hand, we assume $V_0$ is minimized for a large value of \mnu. Thus, these two terms compete, with a minimum at an intermediate value of \mnu, with non-zero value for $V_0$. As the universe expands, the neutrino background dilutes, decreasing the source term, and \mnu\ is driven to larger values.

At the minimum of the effective potential,
\beq
\label{condition}
V'(m_\nu) =n_\nu + V'_0(m_\nu)=0\ .
\eeq
The equation of state of the total   energy density  from this sector is
\beq
w+1 &=& -\frac{\partial \log V}{3\partial \log a} =- \frac{a}{3V}\left(m_\nu \frac{\partial n}{\partial a}+\frac{\partial m_\nu}{\partial a} n_\nu + V'_0(m_\nu) \frac{\partial m_\nu}{\partial a}\right) \\ 
\label{eq:relation}
&=& \frac{m_\nu n_\nu}{V} = 
\frac{\Omega_\nu}{\Omega_\nu + \Omega_{\cal A}}
\\ &=&-\frac{m_\nu V_0'(m_\nu)}{V}\ .
\eeq
These equations already contain a great deal of information. Since the observed equation of state is close to $w=-1$, we know that the energy density in neutrinos alone should be small compared to the energy density in the total dark energy sector.  This corresponds to a fairly flat  potential $V_0(m_\nu)$, either because the potential for \acc\  is rather flat, or because the dependence of $m_\nu$ on \acc\ is steep. Note however,  that the constraints of the potential are not nearly as severe as, \eg,  those in  models  of quintessence, where the ratios of  derivatives of the potential  to the potential itself are required to be smaller than the Hubble scale. 
Good candidate potentials $V_0(m_\nu)\ $ grow as small fractional powers or   logarithms of the neutrino mass.

To the extent that $dw/dz\ $ is small, or, equivalently, that $dw/dn_\nu$ is small, we have the relationship 
\begin{equation}\label{eq:massdensity}
 \mnu  \propto  n_\nu^w\ .
\end{equation}
That is,  the neutrino mass increases   as the neutrino density decreases. For $w\rightarrow -1$, the neutrino mass is simply inversely proportional to the neutrino density.

\subsection{Clustering and constraints}
\label{cluster}
In the previous discussion, we have made the two  assumptions that a) the neutrino energy density is uniformly distributed, and b) we can find the value of the background field  by minimizing its effective potential.  Here we consider the validity of these assumptions, how deviations  affect our scenario, and what additional constraints we need.

We begin by considering the dynamics of the acceleron field \acc. So far we have assumed that  we can treat the neutrino source term  for \acc\ as being spatially constant. In order  that \acc\  does not vary significantly on distances of the order of the inter-neutrino spacing, which, for thermal relic neutrinos, is currently $O(10^{-4}\ \ev)$,  the mass of the  \acc\  particle at the present time must be less than $\sim(10^{-4}\ \ev)$. 
Such a light new particle could also couple  to electrons and nucleons, and  lead to an observable  new force on scales longer than a millimeter---we consider expectations for such couplings and constraints in sec. \ref{acceleron}. Another issue is whether the acceleron field remains at the minimum  of its potential or oscillates. It is self-consistent to assume the acceleron field evolves adiabatically and remains at its minimum provided the time and distance scales over which the neutrino density varies are long compared with the oscillation period.  A lower bound on the mass may be derived from the assumptions that the acceleron has evolved adiabatically since nucleosynthesis, and that the mass has either not changed significantly, or has increased since then, as is the case in the prototypical models we discuss later.  The acceleron mass should then be greater than the Hubble constant at nucleosynthesis, about $10^{-17}\ \ev$.
There are various  ways to explain how such a light field could have settled down to the minimum of its potential prior to nucleosynthesis \cite{Adelberger:2003zx}---as these are all quite model dependent we will not discuss them here\footnote{It is tempting to speculate that oscillations of the acceleron field  could provide the dark matter. If the acceleron potential, averaged over an oscillation,  is flatter than quadratic \cite{Damour:1998cb,Boyle:2001du,Kasuya:2001pr}, acceleron dark matter would  have negative pressure, similar to dark energy. However it would be unstable to clumping  \cite{Damour:1998cb,Boyle:2001du,Kasuya:2001pr,Kasuya:2002zs}. Such clumping would continue until the pressure vanishes, and subsequent evolution would be like normal pressureless matter. Also interesting to consider is the case of a steeper than quadratic potential, leading to positive pressure. One might hope \cite{Anderson:1997un,Comelli:2003cv,Farrar:2003uw,Padmanabhan} that obtaining dark matter and dark energy from the same source would help address the dark matter coincidence problem. However, the discussion of section~\ref{damn} shows that this is not the case. }. 

We now consider the effects of neutrino inhomogeneity. 
Once neutrinos have become sufficiently nonrelativistic, they can cluster. Gravity will pull   some neutrinos into existing dark matter halos. For instance, the authors of  ref.  \cite{Singh:2002de} have studied the gravitational clustering of massive neutrinos in the background of dark matter halos modeled in existing high-resolution numerical studies, and found that significant overdensities can occur for massive neutrinos. Since in our framework, the neutrino mass is not a constant, one cannot directly appropriate the results. However, the vast majority of the clustering happens within a $z$ of one, and a reasonable estimate is to use the neutrino mass at $z=1$ as a benchmark. For instance, for a neutrino of mass $0.6\ \ev$ at $z=1$, an overdensity of $\sim 30$ in the local group is reasonable. Since the cosmological neutrino density has changed by a factor of eight since that time, assuming the neutrino mass is inversely proportional to the local density gives a neutrino mass  in our vicinity today  of $ \sim 0.6 (8/30)\ev \sim 0.15\ \ev$. 

These numbers are rough estimates, but they illustrate an important point.  The ``cosmological'' neutrino mass (outside gravitationally bound systems) would then be  $\CO(5\ \ev)$, which implies $ w \approx - 0.8$.  At the same time, one would measure a mass on Earth of $\mnu \sim 0.15\ \ev$. Without considering the implications of clustering, such a low \mnu\ on Earth would seem in conflict with the equation of state in this model. 
 Precise numerical simulations will be needed in order to determine the proper relationship between the local neutrino mass,  the current cosmological neutrino mass, and cosmological data.

A further possible source of neutrino clustering is the  force between neutrinos mediated by the acceleron.   These effects are quite model dependent, requiring knowledge of the range and strength of the force, and also   numerical study, which is beyond the scope of the present paper. Here we simply note that clustering is suppressed by  several factors, including the short range  of the neutrino force and  the nonnegligible  neutrino velocity.  In any case,
for $w$ near -1, the homogeneity of the dark energy is affected very little by neutrino clustering. Eqns. \ref{eq:relation} and \ref{eq:massdensity} show that the dark energy density  varies with the neutrino density as  $n_\nu^{(1+w)}$.  Neutrino clustering is of interest mainly due to its effects on neutrino mass phenomenology.

\section{Models with a variable mass singlet}
\label{sec:models}
A great deal of phenomenology can be learned directly from model independent features of the MaVaN scenario, but certain questions --- specifically the  dynamics --- require  a specific model. Dirac neutrinos require SM singlet fermions, and Majorana neutrino masses (a dimension five operator in the standard model), often arise from integrating out sterile fermions. A very simple implementation of this scenario is to allow the sterile neutrino mass to vary, causing the mass of the light eigenstate mass to change  accordingly.

As a general framework, we will assume that the active neutrino mass will arise from integrating out a heavier sterile state, whose mass depends on a light scalar field. We take a Lagrangian of the form
\beq
{\cal L} = m_{lr} \nu_l \nu_r +M(\acc) \nu_r \nu_r + h.c.+ \Lambda^4 \log\left(1+ \left| M(\acc)/\mu\right| \right)\ ,
\label{eq:model}
\eeq
where $\nu_r$ is the sterile state, and  will assume 
\beq M(\acc)/\mu\gg1\ .\eeq
At scales below the right handed neutrino mass, we have the effective Lagrangian
\beq
{\cal L} = \frac{m_{lr}^2}{M(\acc)} \nu_l \nu_l + h.c. + \Lambda^4 \log(M(\acc)/\mu)\ .
\eeq
In general, will find $\Lambda \sim 10^{-3}\ \ev$ and naturalness constraints discussed later suggest $\mlr \sim 1\ev$.

From the discussion in section \ref{sec:MaVaNs} all models of this form will have the same equation of state, and the same neutrino mass  for a give neutrino density. However, depending on the form of $M(\acc)$ which we choose, the dynamics can be quite different. We will consider two limiting cases, one in which $M(\acc) = \lambda \acc$ and another where $M(\acc) = M e^{\acc^2/f^2}$. The scalar potential in the former case is very flat, while in the second it is just an ordinary quadratic potential. We will refer to these as the linear model and the exponential model, respectively. Notice that the second case is not unlike what might appear in some models where $\acc$ is a dilaton field for a new force which is the dynamical origin of the
mass term $M(\acc)$. We leave consideration of the possible experimental signals of such a new force coupled to the sterile neutrino sector for future work.

These two models should by no means be taken as the only possibilities. We choose them because they illustrate a wide range of possible dynamics which one might find in this framework. Although it is interesting to consider the the origin of these potentials and couplings, and possible connection with the other fermion masses, and other sectors of the theory, here we leave these issues aside.

\subsection{Quantum corrections and naturalness}
To have an equation of state near $w=-1$ one requires a rather  flat  potential for the neutrino mass. In general, quantum effects  lead to   corrections to the  potential, which are proportional to the second and fourth powers of  the masses of virtual particles. Thus,  from quite model-independent considerations, we can see that the equation of state from theories of variable mass particles is strongly constrained by naturalness issues. For the moment, let us consider a  variable mass particle with mass $\bar m$, and the correction to the potential for $\bar m$.
\FIGURE[t]{
 \centerline{\epsfxsize=3 in \epsfbox{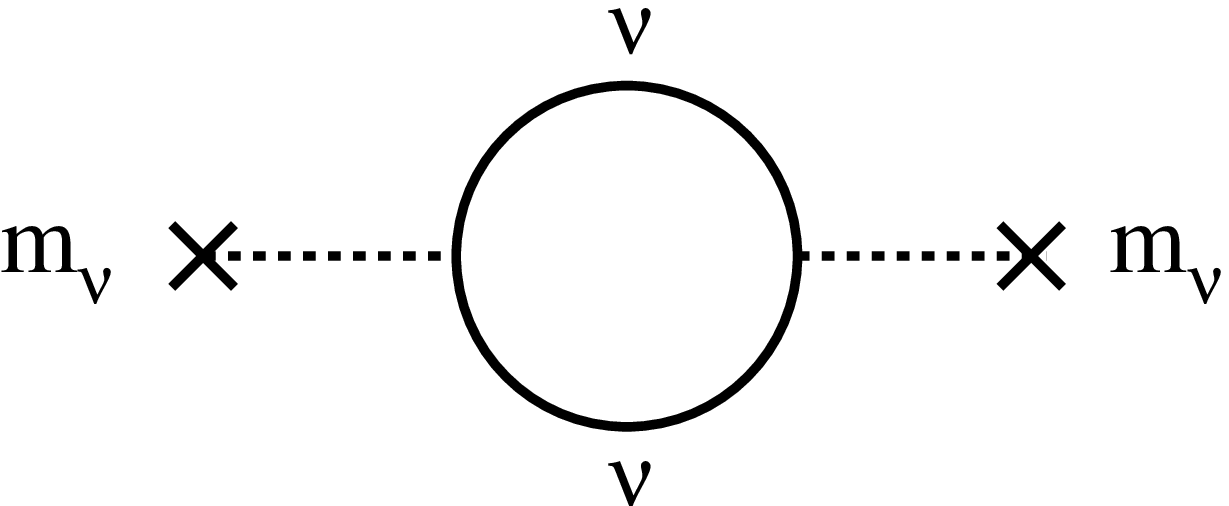}  }
\noindent
\caption{One loop contribution to the \mnu \ potential.}
\label{fig:1loop}
}

At one-loop, we {\em necessarily} have the quadratically divergent contribution to the scalar potential in figure \ref{fig:1loop}. We will throw out the contribution proportional to $\Lambda^4$, and assume that either the true cosmological constant is zero, or that IR dynamics of gravity eliminates its effects. We are left with a term
\beq
\label{quadraticcw}
\delta V_0 \sim \frac{{\bar m}^2 \Lambda^2}{16 \pi^2}\ ,
\eeq
where $\Lambda$ is the cutoff of the theory. 

The  dependence on $\Lambda^2$ indicates that this correction is quadratically sensitive  to the cutoff scale physics, and one might assume  that this short distance physics  for some  reason is such that the contribution  eqn.~\ref{quadraticcw} vanishes, although this is conventionally thought to be unnatural. However, there is also  the cutoff insensitive   correction to the potential
\beq
\delta V_0 = -\frac{1}{32 \pi^2} {\bar m}^4 \log(\bar m/\mu)\ ,\eeq
where $\mu$ is the renormalization scale, which we should take to be of order the value of $\bar m$ today. 

If we want such a variable mass particle to give dark energy with  an equation of state parameter $w$ close to $-1$, we require
\beq
\left| \frac{\delta V_0'(\bar m ) \bar m}{V} \right|< 1\Rightarrow {\bar m}^4\ \lsim \ (10^{-2} \ev)^4.
\eeq

As this argument applies to all particles of varying mass, we see heavy particles are generally unsuitable\footnote{An exception would be if the heavy particle were a member of a nearly degenerate supermultiplet.} but  that the neutrino is  an ideal candidate for contributing to the dark energy. Conventional naturalness implies that the quadratic terms in the potential  are sufficiently small only if the cutoff of the neutrino mass sector is less than of order an \ev. Examples of natural MaVaN models  with a low cutoff for quadratic divergences  include a neutrino mass due to integrating out  a sterile neutrino which has mass near an \ev\  and is either a member of  some highly supersymmetric sector, or is  an extended object,  with a mass and compositeness scale of order an \ev.  In any event, naturalness suggests that new  \ev\  mass states would be detectable at the upcoming MiniBooNE experiment, which can test for neutrino oscillations into sterile states in this mass range.

One might be concerned that this introduces another coincidence into the problem: namely that the present neutrino mass is fairly near the expected cutoff of the theory. However, as total dark energy is set by the cutoff of the theory, this coincidence is equivalent to the fact that we live during the era when the neutrino mass is comparable to the fourth root of the dark energy, and equivalently to the accident that the cube root of the neutrino number density is within a factor of 10 of the fourth root of the dark energy. This is, in turn, directly related to the coincidence of the loosely similar values of  the fourth root of the dark energy and the temperature of the microwave background, or to the fact that the radiation density was comparable to today's dark energy density, 
at a redshift of order 10.
Since the neutrino mass is near the expected cutoff of the theory, we would expect that the present expansion will not last for long. Once the maximal value of the neutrino mass is reached, the scalar field is at its minimum and, since we have assumed vanishing cosmological constant,  accelerated expansion will cease. The observational implications of  this prophesy 
are limited to the expectation of new \ev-scale states with couplings to the neutrino.

\section{Implications for Phenomenology, Cosmology and Astrophysics}
\label{sec:pheno}

In section 2 we showed that if only nonrelativistic neutrinos of uniform number density are present, then their mass
varies as $n_{\nu}^w$.  Here we consider the case where both background neutrinos and relativistic neutrinos are present.  
We will find situations where the extra neutrinos significantly change the mass from the background dominated case and
examine the implications for neutrino mass limits. 

The contribution of neutrinos to the energy density is still given by eqn. (\ref{eq:deltav})
\beq
\delta V = \int \frac{d^3 k}{(2\pi)^3} {\sqrt{k^2 + \mnu^2}} f(k)\ ,
\eeq
but now $f(k)$ includes both background neutrinos and those from local sources.  Since these sources are
nuclear or high temperature ``neutrinostrahlung'' reactions, the additional neutrinos are relativistic
and we can write $f(k) = f_{C{\nu}B}(k) + f_{rel}(k)$.  The local neutrino mass is now given by
\beq
V'(m_\nu) = n_{\nu}^{C{\nu}B} +  \int \frac{d^3 k}{(2\pi)^3} \frac{m_\nu}{\sqrt{k^2
+ \mnu^2}} f_{rel}(k)  +  V'_0(m_\nu)=0 \ .
\eeq

The local neutrino mass will differ from its background dominated value when the second term is larger than
or of the order of $n_{\nu}^{C{\nu}B}$.  An estimate of this can be made by assuming the relativistic
neutrinos all have the same energy $E_{\nu} = k \gg \mnu$.  In this case the above expression becomes
\beq
n_{\nu}^{C{\nu}B} + \frac{ \mnu   }{ E_{\nu}  }n_{\nu}^{rel} + V'_0(m_\nu)=0 \ ,
\label{eq:zeroderiv}
\eeq
where $n_{\nu}^{rel}$ is the density of the relativistic neutrinos as seen in the background frame.

Our approximate measure for finding a non-background mass is to take the ratio of the first two terms in 
(\ref{eq:zeroderiv}), 
{\em 
assuming} that \mnu\ does not change. For practical purposes we shall choose $1\  \ev$ as a value for \mnu\ 
which may be an overestimate in some cases, but since we are seeking an approximate test for interesting 
environments, this is sufficient. Thus we define
\beq 
r \equiv \frac{ 1\ \ev}{ E_{\nu}^{rel} } \frac{n_{\nu}^{rel} }{n_{\nu}^{C{\nu}B} } \ .
\eeq
When this is large, the local neutrino mass differs from its value in a background dominated environment.  
Since $V_0$ is a decreasing function of $m_{\nu}$, larger neutrino densities lower the local mass, and since 
the density cannot be lower than that of the background, neutrinos in a region dominated by the background
have the heaviest mass possible.

We will also take a number density of neutrinos $n_{\nu}^{C{\nu}B} \sim 100\ {\rm cm}^{-3} = 8 \times 10^{-13}$
eV$^3$, which is the expected number for one species of thermally produced neutrino. As we have discussed,
gravitational clustering can raise this value, which will, in turn, result in smaller values of $r$. Since we are
merely interested in isolating those situations which are of possible interest, this is a conservative assumption.
We shall use values of $E_{\nu}^{rel}$ and $n_{\nu}^{rel}$ which are characteristic of the environment under
consideration. Note that the scaling properties of the neutrino mass already described are only applicable for non-relativistic neutrinos. If relativistic neutrinos dominate, the scaling will be considerably different. As this is particularly relevant in the early universe, we will discuss this in detail in section \ref{sec:BBN}.

We will now use this measure to show that all neutrino environments on Earth are background dominated.
 Since $r$ is an order of magnitude estimate, we will pass over the model dependent question of which
neutrino flavors couple to \acc, and use the flavor with the highest density to look for environments
where relativistic neutrinos result in a lower mass.

The most numerous solar neutrinos are the $pp$ neutrinos, with energies of order 0.1 MeV and a flux on earth
of $6 \times 10^{10}$ cm$^{-2}$s$^{-1}$ \cite{Bahcall:1989ks}.  Their number density is thus 2 cm$^{-3} = 2
\times 10^{-14}$ eV$^3$, so $r \sim 10^{-7}$ and the background neutrinos dominate solar neutrinos on earth.  
In the solar core $r$ is larger, but we will discuss this in sections (\ref{sect:intenv}) and 
(\ref{sect:solar}).

In reactor cores, the neutrino density can reach $\sim 10^3$ cm$^{-3}$ $\sim 10^{-11}$ eV$^3$
\cite{Barbeau:2002fg}, and the energies are again of order 0.1 MeV, so there again $r$ is small ($\sim
10^{-4}$).

In the most intense neutrino beams like those at K2K \cite{Ahn:2001cq} and MiniBooNE \cite{Link:2002rz}, proton
pulses collide with a target to produce pions that decay into GeV muon neutrinos.  One can consider the limit in which a neutrino is produced for each proton on target. With roughly $6 \times 10^{12}$ pot (protons on target) in $1.1\ \mu s$, with a source of cross section $\sim 5\ {\rm cm}^2$, one yields an upper limit of $4 \times 10^7 \ \nu \ {\rm cm}^{-3}$. For \gev\ neutrinos, this still yields an upper limit on $r$ less than one.

Neutrinos from other sources such as the Earths radioactivity \cite{Rothschild:1997dd} and cosmic rays are too
diffuse to be important.  We note that nuclear fireballs might briefly produce neutrino densities large enough to 
change their mass, but no physics has been done with fireball neutrinos and we do not propose it should be.

We conclude that the local background neutrinos dominate in all earth based experiments and that the tritium 
beta decay limit \cite{Lobashev:1999tp} gives the most stringent upper bound of $\sim$ eV for the background 
neutrino mass.

\subsection{Interesting Environments}
\label{sect:intenv}

For neutrino masses to vary significantly, we need either a high neutrino density relative to the background, or if
few high energy neutrinos are present, then significant variation in the background itself.  As discussed in section
(\ref{cluster}), we will not consider background variations in depth, but we will now turn to several high
density cases.  They are in the early universe, supernovae, and stellar cores.

Redshifting $100$ $\nu$ cm$^{-3}$ back to the BBN era with a $z$ of $10^{14}$, we find neutrino number
densities of $10^{30}$ eV$^3$ and extremely light active neutrinos ($r \sim 10^{36}$).

In core collapse supernovae, the early neutrino number density can reach 10$^{35}$ cm$^{-3}$ ($10^{21}$ eV$^3$)  
\cite{Raffelt:1996wa}, which with order MeV energy gives an $r$ of $\sim 10^{27}$, certainly enough to lighten our
active neutrinos.  A separate issue from simply lowering the local neutrino mass is whether \acc\ emission changes
the rate of cooling of the young neutron star, but we show in section (\ref{sect:super}) that the effect is small
over broad regions of parameter space.

While the above two environments are clearly dominated by relativistic neutrinos, stellar cores are a
marginal case.  The $pp$ neutrinos that failed to dominate the background on earth are denser in the solar core.  
Within roughly 0.1 $R_{\odot}$ \cite{Bahcall:1989ks}, their density is $\sim$ (1 AU/0.1 $R_{\odot}$)$^2 \times 2 
\times 10^{-14}$ eV$^3 = 9 \times 10^{-8}$ eV$^3$, and an energy of 0.1 MeV gives them an $r$ of order 1.  Higher 
energy neutrinos such as those from $^8$B reactions have a negligible $r$ due to both their lower number density and 
higher energy.  Thermally produced neutrinos from pair reactions such as $\gamma e \rightarrow e \nu \nu$ have a lower 
energy but are also more diffuse and have a negligible $r \sim 10^{-3}$ \cite{Haxton:2000xb}.  An $r>1$ for any of the components will affect the masses of all neutrinos, of course, so the pp neutrinos may have an impact on the MSW properties in the sun. We will discuss this in greater detail in sections \ref{sect:solar}.

Before discussing these cases further, we note two other environments in which neutrino masses could vary
significantly.  White dwarves cool largely by thermally produced neutrinos and can easily have neutrino densities of 
order $10^{13}$ cm$^{-3}$ which gives a large $r \sim 10^7$ (see Appendix C of \cite{Raffelt:1996wa}).  Also, large 
stars such as red giants can produce more neutrinos from smaller regions than in the sun \cite{Raffelt:1996wa}.  As 
the solar case is already marginal, we may find reduced neutrino masses in the cores of larger stars.  However, since 
no observations have been made with neutrinos from these sources we will not treat them further here.

Table 1 gives a summary of environments from which neutrino physics has been extracted, with $r \ge 1$ showing 
those in which the neutrino mass is lowered.

\begin{table}[t]
\begin{center}
\begin{tabular}{|c|c|c|c|} \hline
                & $n_{\nu}$ eV$^3$ & $E_{\nu}$ & r \\ \hline
BBN             & $10^{30}$ & $\sim$ MeV & $10^{36}$ \\ \hline
SN cores        & $10^{21}$ & $\sim$ MeV & $10^{27}$ \\ \hline
solar core      & $\sim 10^{-7}$ & 0.1 MeV & 1 \\ \hline
reactor cores   & $10^{-11}$ & 0.1 MeV & $10^{-4}$  \\ \hline
beams           & $< 3\times 10^{-7}$ & GeV & $<10^{-2}$ \\ \hline \end{tabular}


\end{center}
\caption{$r$ values for various environments. $r \ge 1$ implies that neutrino mass parameters could be 
modified.}
\end{table}

We now consider the BBN, SN, and solar cases in slightly more detail.

\subsection{Limits on sterile neutrino and acceleron production from BBN}
\label{sec:BBN}
Under ordinary circumstances, a sterile neutrino that has both a) appreciable mixing, and b) sufficiently small mass,  will come into thermal equilibrium prior to nucleosynthesis. This can have serious consequences for BBN, making models with sterile neutrinos very constrained, cosmologically. Here, we have not only a sterile neutrino, but also the possibility of producing light $\acc$ quanta, so one must take care that BBN is not radically changed.

Concerning the sterile neutrino, the key point is that while parameters may be interesting today (\ie, 
accessible at the MiniBooNE experiment), these are effectively unrelated to the values at times prior to 
nucleosynthesis. It is important to note that the relationship $\mnu \propto n_\nu^w$ holds only for 
non-relativistic neutrinos. For relativistic neutrinos, we need to minimize the potential with the relativistic form for the scalar neutrino density, namely
\beq
\frac{\mnu n_\nu}{T} + V'_0(\mnu)=0\ .
\eeq
Following the same procedure as we did for non-relativistic neutrinos, we arrive at
\beq
w+1 = \frac{\mnu^2 n_\nu}{3 T V}= - \frac{\mnu V_0'(\mnu)}{3 V}\ .
\eeq
Now, assuming a slow variation in $w$, we have
\beq
\mnu \propto a^{(3w+1)/2}\ ,
\eeq
which is considerably slower than in the non-relativistic case. 

In the microscopic models presented, it was a sterile neutrino whose mass was varying, and hence, for a relativistic system, the sterile neutrino mass will vary differently in the hot early universe compared with later.  
In particular, if we consider the model in Eq. \ref{eq:model}, then for relativistic neutrinos, the expectation value of for $M(\acc)$ (the sterile neutrino mass) is
\beq
M(\acc) \approx \frac{T m_{lr}^2}{\Lambda^2}\ .
\eeq
Since $m_{lr}> \mnu > \Lambda$ the sterile
 state should be above the temperature scale and should not affect the light element abundances.

We now turn our attention to the acceleron field which is setting the mass of the right handed neutrino.
We certainly expect acceleron  particles to be light (no heavier than $10^{-4}\ \ev$ at the present era), and so they can be potentially thermalized in the early universe as well.
\FIGURE[t]{
 \centerline{a)\ \ \epsfxsize=2.5 in \epsfbox{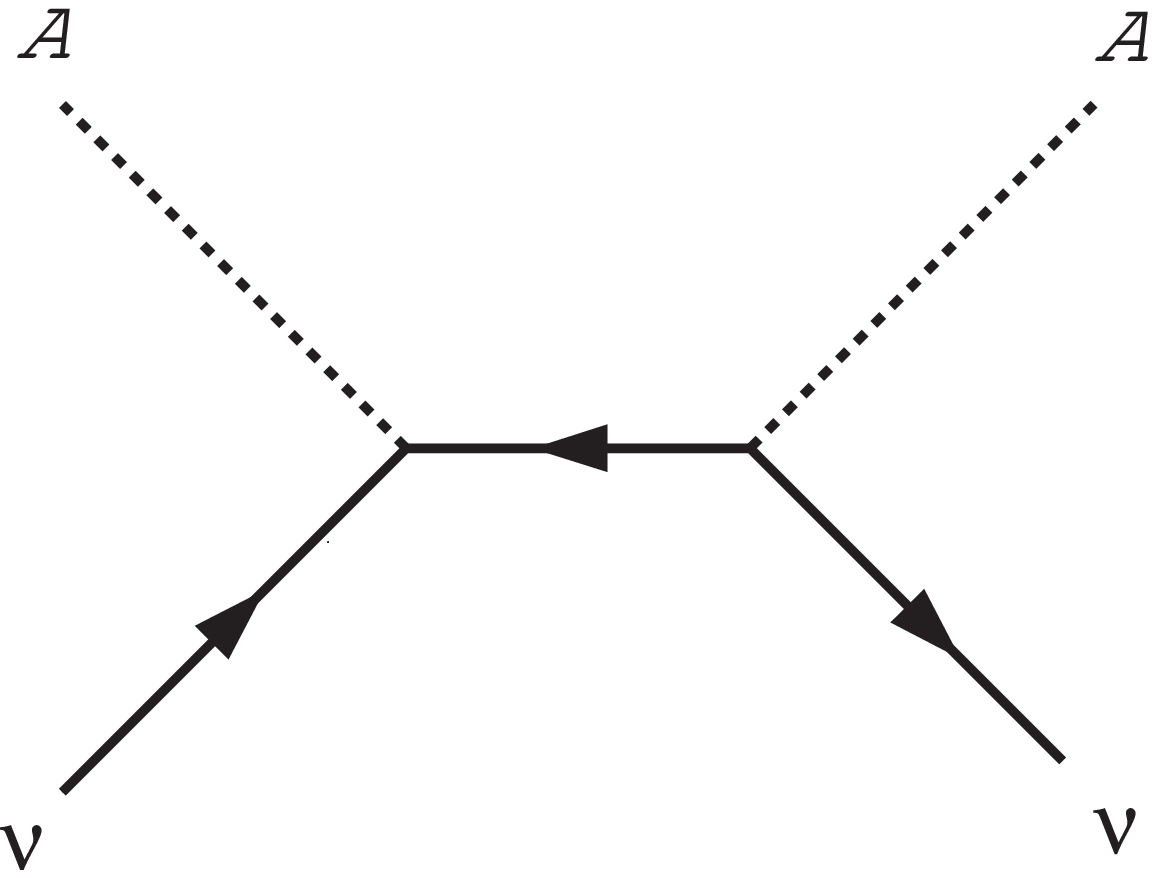} \hskip 1in b) \ \epsfxsize=1.75 in \epsfbox{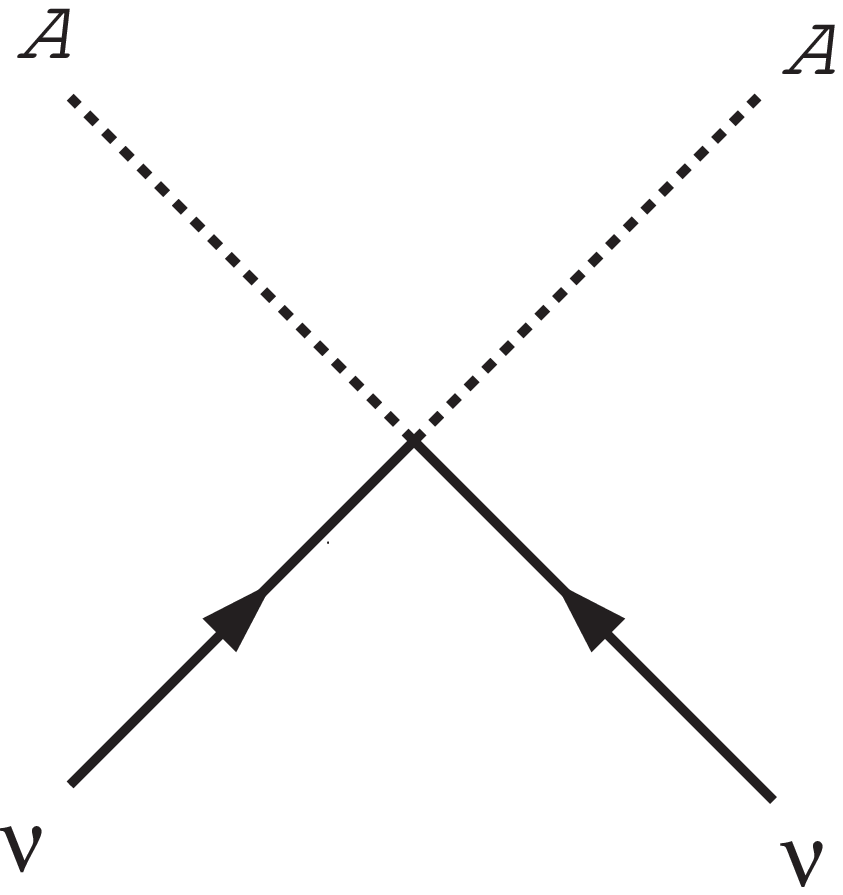} }
\noindent
\caption{Neutrino annihilation diagrams.}
\label{fig:annihilate}
}

Through the diagram of figure \ref{fig:annihilate}a, we expect an annihilation rate in the early universe
\beq
\frac{1}{\tau} \sim \left(\frac{\partial \mnu}{\partial \acc}\right)^4 T\ ,
\eeq
while through the diagram of figure \ref{fig:annihilate}b, we expect a rate
\beq
\frac{1}{\tau} \sim \left(\frac{\partial^2 \mnu}{\partial \acc^2} \right)^2 T^3\ .
\eeq
In addition, there can be \acc\  production through \acc-strahlung following a weak interaction, which has a production rate
\beq
G_F^2 \left(\frac{\partial \mnu}{\partial \acc} \right)^2 T^5\ ,
\eeq
so the requirement that these particles are not in thermal equilibrium gives us
\beq
\nonumber \left(\frac{\partial \mnu}{\partial \acc}\right)^4 < \frac{T}{\mpl} \sim 10^{-22}\ ,\\
\left(\frac{\partial^2 \mnu}{\partial \acc^2}\right)^2 < \frac{1}{T\mpl} \sim 10^{-34} \ev^{-2}\ ,\\
\nonumber \left(\frac{\partial \mnu}{\partial \acc}\right)^2 \lsim 1\ ,
\eeq
where the final relationship comes from evaluating these expressions at the era of BBN.

Let us consider our two example models. In the linear model, we have during nucleosynthesis
\beq
\vev{\acc} \approx { T m_{lr}^2 \over \lambda \Lambda^2}\ ,
\eeq
hence
\beq
\left|\frac{\partial \mnu}{\partial \acc} \right|= \frac{ \mlr^2}{\lambda \acc^2} = \frac{ \lambda \Lambda^4 }{T^2 m_{lr}^2} \lsim 10^{-16}\ ,\\ \nonumber
\frac{\partial^2 \mnu}{\partial \acc^2} = \frac{2 \lambda^2 \Lambda^6}{\mlr^4 T^3} \lsim 10^{-22} \ev^{-1} \ ,
\eeq
hence neither contributions to the production of $\acc$ will thermalize it.

In the exponential model we have
\beq
\frac{\partial \mnu}{\partial \acc} = \frac{-2 \Lambda^2 \acc }{T f^2} \ ,\\
\nonumber \frac{\partial^2 \mnu}{\partial \acc^2} = \frac{\Lambda^2}{T}\left(\frac{4\acc^2 -2 f^2}{f^4}\right) \ .
\eeq
hence to avoid thermalization, one should have $f \gsim 450\ ev$.

Interestingly, because the light neutrino mass eigenstate is becoming increasingly sterile at late times, the production of acceleron quanta can happen at late times. Should this occur when the neutrinos are non-relativistic, their number density would become exponentially suppressed, which would be problematic even for logarithmically flat potentials. Because the particles are Majorana fermions, there is an additional velocity suppression of their cross sections when they are non-relativistic, and hence the annihilation rate of neutrinos drops faster than the Hubble rate. Hence, we need only check that they are not in thermal equilibrium at the scale when they go non-relativistic. In the models we are considering, this is when 
\beq
T\approx \mnu \Rightarrow \Lambda \approx T
\eeq
Since $\Lambda \sim 10^{-3} \ev$, this occurred at $z\simeq 10$ at which time the Hubble scale was different by a factor of roughly ten.

We begin by considering the linear model. At the non-relativistic transition, the annihilation rate goes as from both diagrams goes as
\beq
\frac{1}{\tau} \sim \frac{\lambda^4 \Lambda^9}{\mlr^8}
\eeq
And requiring this to be less than the Hubble scale requires
\beq
\frac{\mlr}{\sqrt{\lambda}} \gsim 4\ \ev
\eeq
So for $\lambda \sim 1/100$ the scenarios discussed should be safe. For larger values of $\lambda$ and lower values of \mlr\ a more rigorous analysis should be performed.

In the exponential model, at the transition, we have (taking $\vev{A} \sim f$, which is natural if the sterile neutrino mass is to be of order $\ev$.)
\beq
\frac{1}{\tau} \sim \frac{\Lambda^5}{f^4} 
\eeq
Requiring this to be less than Hubble at $z\simeq 10$, we yield 
\beq
f\gsim 10\ \kev \ ,
\eeq
which is more restrictive than previous limits.

\subsection{Supernovae}
\label{sect:super}

Supernovae are interesting to us for several reasons.  In contrast to neutrinos from white dwarves and red giants, 
neutrinos from a core collapse supernova have been observed \cite{Alekseev:1987km,Hirata:1987hu,Bionta:1987qt}, and we 
must check that \acc\ emission does not cool a protoneutron star too quickly to alter the observed spectrum of 
neutrinos emitted in the first 10 seconds.  Further, if r-process nucleosynthesis occurs in the envelopes of
core collapse supernovae, it will be sensitive to the flavor composition of the neutrino wind emanating
from the newly formed neutron star.

We can quickly check that \acc\ emission is not significant.  Neutrinos inside the early neutrino sphere have 
energies of order 10 MeV, and we need emission rates of less than 1/10 sec$^{-1}$.  This requires only slightly 
stronger limits as those from BBN:

\beq
\nonumber \left(\frac{\partial \mnu}{\partial \acc}\right)^4 \lsim 10^{-23}\ ,\\
\left(\frac{\partial^2 \mnu}{\partial \acc^2}\right)^2 \lsim 10^{-37} \ev^{-2}\ ,\\
\nonumber \left(\frac{\partial \mnu}{\partial \acc}\right)^2 \lsim 10^{-12} \ .
\eeq

These are still easily satisfied by the linear model.  The \acc-strahlung limit is itself tighter than the BBN 
limit, but it is again redundant given the limit from $\nu \nu \rightarrow \acc \acc $.  Thus \acc\ emission does not 
speed the early cooling of supernovae.

A cooler neutrino sphere persists for roughly 10-100 years \cite{Raffelt:1996wa}, during which time neutrino
cooling dominates in the standard picture.  This again may be altered by \acc\ emission in our case, and the longer 
timescale sets somewhat tighter limits than those imposed in the first 10 seconds.  However, since cooling rates of 
young neutron stars have not been accurately measured, due to the difficulty in separating X-rays from the stars 
surface with other sources \cite{Pavlov:2002ec}, we will not pursue this further here.

The r-process is thought to occur in the envelopes surrounding core collapse supernovae, where neutron
capture onto heavy nuclei proceeds more rapidly than the beta decays of these nuclei.  The process is
sensitive to the $\nu_e$ flux from the core, since for example the $\nu_e$'s can turn the free neutrons into 
protons, reducing the neutrons available for the r-process.  Since these protons end up inside $\alpha$ particles 
this problem is known as the ``alpha effect'' \cite{Meyer:1998sn, Fuller:1995ih}.  Both active-active 
oscillations \cite{Qian:1993dg, Sigl:1995hc} and active-sterile oscillations 
\cite{Peltoniemi:1995dp,Caldwell:1999zk,McLaughlin:1999pd} have been suggested
as important mechanisms for both the r-process and shock dynamics, and it is possible that active-sterile
mixing with a $\delta m^2$ of 0.1 - 100 eV$^2$ and matter enhancement might solve the above ``alpha problem''
\cite{Balantekin:2003ip}.

In our variable mass singlet model, we have a sterile state whose mass would be $\sim$ MeV in the core, that becomes
progressively lighter as it propagates outwards into the less dense envelope, and an active state that get
progressively heavier and more sterile.  Should MiniBooNE see a positive result, its implications within the MaVaN
scenario for the r-process could be significantly different and would need to be carefully analyzed.

\subsection{Solar neutrinos}
\label{sect:solar}

Our estimate for $pp$ neutrinos in the solar core shows that a detailed calculation is needed in the event that \acc\ 
couples to $\nu_e$.  If the electron neutrino mass is modified, this could affect our interpretation of the KamLAND and SNO results\footnote{A similar effect has been studied in the context of ``neutrino
clouds''\cite{Stephenson:1997cz}, although the effect is due to clustered primordial neutrinos, not pp neutrinos.}.  For neutrinos with constant mass, $\nu_e$s are produced in the sun's core and oscillate into 
$\nu_{\mu}$s and $\nu_\tau$s via the MSW effect whilst still inside the sun.  In our scenario, background neutrinos 
dominate the acceleron potential until we are sufficiently near the core of the sun, at which point the mass of the electron neutrino could be 
lowered depending on it's energy.  There is the possibility that the neutrino mass matrix would be altered near the solar core, modifying the interpretation of the solar neutrino results, but this is model dependent and the detailed consequences are yet to be worked out.

\subsection{Baryogenesis}
There are many baryogenesis scenarios which constrain neutrino masses---see for example refs. \cite{Fukugita:1986hr,Nelson:1990ir,Harvey:1990qw,Fischler:1991gn,Buchmuller:2000as,Davidson:2002qv,Dolgov:2002wy,Hasegawa:2003vh}. Majorana neutrino masses violate lepton number (L), and, since high temperature nonperturbative  electroweak processes violate baryon number (B) but conserve B-L,   an upper bound on Majorana neutrino masses, which depends on the baryogenesis scale, can be derived.  An attractive baryogenesis mechanism which constrains both the sterile and the light neutrino masses is to use CP violation in the decays of heavy sterile neutrinos to create L, which is subsequently partially converted in B by nonperturbative electroweak interactions. The constraints obtained on neutrino masses from baryogenesis scenarios all assume constant neutrino mass and will clearly  have to be reevaluated in the MaVaN scenario. In the models of sec.~\ref{sec:models}, early on the sterile neutrinos were much heavier and the active neutrinos much lighter, and so the constraints from baryon number washout on neutrino masses become very weak.

\section{Other Roles for the Acceleron?}
\label{acceleron}
\subsection{Long Range Forces}
\label{longrange}
Could the acceleron mediate forces between other particles than neutrinos? There are strong limits on new forces for between nucleons and/or electrons at scales longer than 100 microns \cite{Adelberger:2003zx}. For instance, the operators
\beq
{m_e}(\acc) \bar e e\ ,\qquad{m_d}(\acc)\bar d d\ 
,\qquad {m_u}(\acc)\bar u u  
\ ,\qquad \frac{1}{g_s^2(\acc)}G_{\mu\nu}^a G^{a,\mu\nu}\ . 
\label{eq:ops} 
\eeq 
could lead to violations of the $1/r^2$ force law and of the weak equivalence principle, on scales longer than a millimeter, if $m_e',m_d',m_u', $ or $g_s'\ne0$. The Yukawa force mediated by, \eg, the operator ${m_e}(\acc) \bar e e\ ,$ is proportional to ${m_e'}^2$, and has range $1/m_\acc$. 

Standard Model quantum corrections will necessarily  induce acceleron dependence into such operators, but with negligible coefficients, suppressed by loop factors and by  a factor at least as small as 
\be  G_F  \frac{d m^2_\nu}{ d \acc} \ .\ee  In the explicit models of sec. \ref{sec:models} the minimum natural size of  any such acceleron mediated force is too small to give observable deviations from the inverse square law or the weak equivalence principle\footnote{Although not interesting for long range forces, the coupling of the acceleron to the electron which is induced by quantum corrections can be large enough to induce phenomenologically important changes in the masses of neutrinos propagating through electron  dense matter such as the core of the  sun or solids on the earth. This effect differs from the usual matter effect \cite{wolfenstein} in being the same for neutrinos and anti neutrinos and in not being neutrino energy dependent. A phenomenological study of such effect is underway \cite{us}.}.  

Although we lack understanding of the role of \acc\ in the fundamental theory,  we can postulate that it could be a modulus for other parameters of the standard model besides neutrino masses, and test for this by searching for violations of the weak equivalence principle and of the inverse square law at distance scales longer than a millimeter. The observation or nonobservation of such effects  constrain the mass and couplings of the acceleron, and guide model building towards its fundamental origin.

\subsection{Varying $\alpha$}
\label{varyingalpha}

Besides leading to new forces,  since the value of \acc\ depends on  $s_\nu$, nontrivial \acc\  dependence of the operators \ref{eq:ops} implies that ``constants'' of the  standard model depend nontrivially on the scalar neutrino density.  This is different from typical treatments of varying constants, which mainly consider temporal variations (for  reviews see ref.~\cite{Sisterna:1990et,Cowie:1995sz,Uzan:2002vq}).
It is particularly interesting to consider the operator
\be\frac{1}{e^2(\acc)}F_{\mu\nu} F^{\mu\nu}\ ,\ee
in light of the recent astrophysical evidence from quasar absorption lines  that $\alpha$ may be variable \cite{Webb:2000mn,Murphy:2003hw}.

Such dependence is highly constrained by BBN, since $s_\nu$ was larger then by a factor of at least $10^{13}$ . The success of BBN implies that the constants of nature have changed little since. One might think that the BBN implies that no interesting observable dependence of $\alpha$ or other constants on $s_\nu$ is possible. However, if the dependence of $\alpha$ on $s_\nu$ is sufficiently flat, \eg\ logarithmic,  then, given that  $s_\nu$ varies steeply with redshift and is likely to vary with, \eg\ the matter density  due to gravitational neutrino clustering, it is possible to reconcile the observation of astrophysically varying  $\alpha$ with BBN.

The hypothesis that $\alpha$ depends on $s_\nu$ explains a troubling aspect of the recent claims. 
It is calculated that the presence  2 Gyr ago of a natural nuclear reactor on
Earth implies that $\alpha$ cannot have fractionally changed by more than  $\sim
10^{-7}$ since then\cite{Damour:1996zw}. Ref. \cite{Webb:2000mn} interprets its
data as evidence that $\alpha$ has increased by a fraction $\sim 10^{-5}$ since
redshift of greater than $\sim 0.5$. If one assumes linear variation of alpha
with time, the results of refs.  \cite{Webb:2000mn}  and \cite{Damour:1996zw}
are inconsistent. However, the local value of $s_\nu$ should have changed little
since the formation of the galaxy, but could be quite different outside our
galaxy or galaxy cluster, and so the terrestrial value of $\alpha$ could be
essentially   constant while the value of $\alpha$ outside the galaxy could be
slightly varying. To test this, one could look for dependence of $\alpha$ on, \eg,
the  matter density, rather than on redshift. Similar effects have been studied
in some mixed quintessence/varying $\alpha$ models \cite{Mota:2003tc,Mota:2003tm}.

\subsection{MaVaNs and High Energy Cosmic Rays from the Z-Burst Scenario}
One of the most puzzling questions in high-energy astrophysics is the nature of the highest energy cosmic rays \cite{Takeda:1998ps,Bird:1993yi,Bird:1994wp,Bird:1995uy,Lawrence:1991cc,Ave:2000nd,Kieda:1999ig,Efimov:1990rk}. The expected Greisen-Zatsepin-Kuzmin (GZK) cutoff \cite{Greisen:1966jv,Zatsepin:1966jv} in their energy has not been observed \cite{DeMarco:2003ib}, suggesting the possibility of an exotic origin of the highest energy cosmic rays.

One possibility is the Z-Burst scenario \cite{Weiler:1982qy} in which ultra-high energy neutrinos resonantly annihilate with thermal relic neutrinos. The possible success of this model depends strongly on the neutrino mass, both to yield cosmic rays of the proper energy, and to induce sufficient clustering to amplify the sources. However, the 68 \% preferred mass range of $0.08\ \ev - 1.30\ \ev$ (extragalactic source for UHE $\nu$'s) or $2.1\ \ev - 6.7\ \ev$ (halo source) \cite{Fodor:2002qf} is significantly different from the measured atmospheric neutrino splitting. If all the neutrinos are degenerate, much of the parameter space is in conflict with limits on the neutrino mass from structure formation.

In the MaVaN scenario, there are two significant differences from ordinary
neutrinos\footnote{The effect of neutrino clustering has also been studied in
  the context of neutrino clouds \cite{Stephenson:1998qj} where considerable changes to the
  Z-burst scenario also occur \cite{McKellar:2001hk}. The physics is quite different, however, as the
  changes are essentially due to the large Fermi momenta of the neutrinos
  inside the neutrino cloud.}. The first is what has already been stated: the
cosmological limits on neutrino masses simply do not apply here. Hence the
entire parameter space is opened up. The second, arises from the different
density of neutrinos in gravitationally bound systems. As we have stated the
measured value of neutrino mass on the Earth need not be the same as in other
systems, nor does the mass in the interior of a cluster need to equal the mass
near the exterior. This would in principle require an integration over a continuum of mass values, rather than a particular value, as in the fixed mass theory. As a consequence, the parameter space for the Z-burst
scenario as it relates to terrestrial measurements of neutrino mass is opened considerably. To
properly understand this, a careful analysis of neutrino clustering along the
lines of \cite{Singh:2002de} would be essential. 

\section{Outlook and Smoking Guns}
Inspired by the similarity between the  the dark energy and neutrino mass 
scales, we have proposed a link between the dark energy and the neutrino 
mass, which implies that the neutrino mass is approximately inversely 
proportional to the scalar neutrino density. The neutrino mass varies in 
such a way that the  total energy density in cosmic neutrinos is always 
similar to the dark energy density.
The fact that   the neutrino mass today is not very different from the 
fourth root on the dark energy is related to  the coincidence that we live 
during the epoch when the microwave background temperature is within an 
order of magnitude of the fourth root of the dark energy. 

 To test the scenario terrestrially requires an extremely intense source of 
neutrinos---intense enough so that the scalar density, which is suppressed 
by $m_\nu/E_\nu$, is  larger than the scalar density of the cosmological 
background neutrinos. To date, the only sufficiently intense terrestrial 
source is a nuclear bomb, in which  we see no way  to   test for neutrino 
masses. There are however several possible future observations which could 
provide distinctive evidence for  MaVaNs. One possibility is that the 
upcoming MiniBooNE experiment \cite{McGregor:2003ds} could provide 
evidence for one or more sterile neutrinos. Since, assuming fixed neutrino 
masses,  the mass and mixings of such detectable  sterile neutrinos  would 
likely be problematic for nucleosynthesis, this alone would suggest that 
the neutrino mass should vary. For solutions to the questions relating to sterile neutrinos in BBN, in particular those involving primordial lepton number or additional fields see \cite{Dolgov:2002wy,Babu:1991at,Foot:1995bm,Bell:1998sr,Bento:2001xi,Abazajian:2002qx,Abazajian:2002bj,Dolgov:2003sg}.
Another possibility is that cosmological or 
supernova constraints on neutrino masses could eventually conflict with 
the neutrino masses derived from observations within our solar 
system--again implying that the neutrino masses should vary. Finally, an  acceleron coupling to the electron or nucleon could lead to an observable   matter density dependence of neutrino oscillations which differs from the usual matter effect\cite{us}. 

There are two model independent predictions of MaVaN dark energy. One 
prediction is the relation of ${\rm eq.}$ \ref{eq:relation} between the dark 
energy equation of state parameter and the cosmological neutrino mass. To 
test this, one needs to precisely measure the equation of state parameter. 
This will determine the cosmological neutrino mass today. To obtain a 
prediction for the local neutrino mass, which depends on the neutrino 
density in our vicinity, one  needs to use detailed numerical modelling to 
study neutrino clustering in our galaxy. The second prediction is that if  
all neutrino masses  vary in the same way,  neutrino masses  at high 
redshift  scale like $(1+z)^{3w}$, and so high redshift cosmological  data 
should show no evidence for neutrino mass.

Still requiring much further study are the consequences of MaVaNs for 
neutrino oscillation experiments, solar neutrino physics, supernovae, 
neutrino clustering, structure formation, baryogenesis,  and high energy 
cosmic rays. Also interesting is the cosmology of and the origin of the 
acceleron, the scalar field which mediates the dark force between 
neutrinos. Searches for deviations from the equivalence principle and the 
$1/r^2$ force law at scales of order a millimeter or longer could shed 
some  light on the acceleron, as could variation of other `constants' with 
the neutrino density.

\bigskip\noindent{\bf Acknowledgments}
The authors would like to thank Lawrence Hall, Craig Hogan, Tom Quinn, Chris Stubbs, Julianne Dalcanton and the entire particle-cosmo lunch group at the University of Washington for useful conversations. This work was partially supported by the DOE under contract DE-FGO3-96-ER40956 and the University of Washington Royalty Research Fund. RF was partially supported by NSF grant AST-0098557.

\vskip 0.15in
\bibliography{MaVaNs}
\bibliographystyle{JHEP}
\end{document}